\documentclass[garamond]{acdis}
\usepackage{algorithm,algpseudocode,tabularx}
\usepackage{placeins}
\usepackage{comment}
\usepackage{todonotes}
\author{Tuna Erdo\u{g}an, Shi-Yuan Wang, Shang-Jen Su, Matthieu Bloch}
\title{Joint Communication and Sensing with Bipartite Entanglement over Bosonic Channels}
\rightnote{}
\leftnote{}

\newcounter{relctr}
\everydisplay\expandafter{\the\everydisplay\setcounter{relctr}{0}}

\AtBeginDocument{} 
\definecolor{navyblue}{RGB}{0, 0, 128}
\hypersetup{
  colorlinks=true,
  linkcolor=navyblue,
  citecolor=navyblue,
  filecolor=navyblue,
  urlcolor=navyblue
}
\usepackage{tikz}
\usepackage{pgf,pgfplots}
\pgfplotsset{compat=newest}
\usepgfplotslibrary{groupplots}
\usepgfplotslibrary{dateplot}
\usetikzlibrary{arrows,arrows.meta,automata,shapes,calc,intersections}
\definecolor{navyblue}{RGB}{0, 0, 128}
\hypersetup{
  colorlinks=true,
  linkcolor=navyblue,
  citecolor=navyblue,
  filecolor=navyblue,
  urlcolor=navyblue
}

\begin{document}
\maketitle
\thispagestyle{fancy}
\allowbreak

\begin{abstract}
  We consider a joint communication and sensing problem over an optical link in which a low-power transmitter simultaneously communicates with a receiver and identifies the range of a defect producing a backscattered signal. We model the system as a lossy thermal-noise bosonic channel, in which the target location, modeled as a beamsplitter, affects the timing of the backscattered signal. Motivated by the envisioned deployment of entanglement-enabled quantum networks, we allow the transmitter to exploit shared entanglement to assist both sensing and communication. Since entanglement is known to enhance sensing, as demonstrated in \ac{QI}, and to increase communication rates through entanglement-assisted communication, the transmitter faces a trade-off in allocating its entanglement resources between the two tasks. Our main result is a characterization of these trade-offs in the form of an achievable rate/error-exponent region, which can outperform time-sharing and demonstrates a quantum advantage.
\end{abstract}

\section{Introduction}
\label{sec:intro}

Although the vision of the Quantum Internet as an ubiquitous network connecting distant quantum computers and sensors is still decades away, there have been recent significant advances towards a first-generation Quantum Internet in the form of an entanglement-sharing network~\cite{Yin2017}. In particular, we now have a deeper understanding of how to efficiently distribute entanglement over long distances by appropriately processing shared entanglement between neighboring nodes~\cite{pant2019routing}. A key aspect of many recent works, however, is to treat entanglement as a resource to be consumed without considering the nature of the tasks consuming entanglement. As long as entanglement-sharing remains costly, efficient use of entanglement is likely to require a careful allocation of entanglement resources across tasks.

Our objective is to develop preliminary insights into the resulting performance trade-offs that would result from this operation by formulating an exemplar combining two entanglement-assisted tasks: sensing and communication. This exemplar is motivated by recent experimental results reporting joint communication and sensing in optical links~\cite{He2023Integrated}, as well as the usefulness of entanglement to improve detection in quantum illumination~\cite{Wilde2017} and increase communication rates over bosonic channels~\cite{Bennett2002}. This model extends our prior work on joint communication and sensing that only considered finite-dimensional Hilbert spaces and did not include entanglement assistance~\cite{Wang2022Joint}, and extends recent independently obtained results on joint communication and sensing over lossy bosonic channels without entanglement~\cite{MunarVallespir2024Joint}. The problem of joint communication and sensing has been investigated in classical information theory under various models: capacity/distortion/cost trade-offs for \ac{iid} channel states~\cite{Ahmadipour2024}, rate/error-exponent trade-offs for binary channel states~\cite{Joudeh2022} and rate/error-exponent tradeoffs under monostatic and bistatic detection models~\cite{Chang2022,Chang2023}.

We consider here a lossy thermal noise bosonic channel in which a transmitter simultaneously wishes to communicate with a receiver and range the location of the target which is modeled as a beamsplitter~\cite{Zhuang2021}. The operation of the transmitter is assisted by an entangled state from a \ac{TMSV} source~\cite{Weedbrook2012} consisting of a signal and idler pair. The transmitter encodes information on the signal but is faced with a dilemma with respect to the the idler: it can transmit the idler to the receiver, i.e., through an entanglement sharing network, to enable entanglement-assisted communication or it can keep the idler, i.e, in a local quantum memory, to enable quantum illumination. We provide insights into the incurred performance trade-offs in the form of an achievable communication rate/sensing exponent.

The remainder of the paper is organized as follows. We formally introduce the notation and joint communication and sensing model with bipartite entanglement resource in Section~\ref{sec:not_and_model}. We present our main results regarding the joint communication and sensing trade-off in Section~\ref{sec:results} and defer proofs to Section~\ref{sec:proofs} to streamline presentation.

\section{Notation and System Model}
\label{sec:not_and_model}
\subsection{Notation}
\label{sec:notation}

Let $\bbR_+$ and $\bbN^*$ denote the sets of non-negative real numbers and positive integers, respectively. For any set $\calX$ and $n\in\bbN^*$, a sequence of $n$ elements is denoted by $x^n=(x_1,\dots,x_n)\in\calX^n$. For any $n$ and $k$ such that $n\leq k$, $x^{n:k}\eqdef(x_n,\dots,x_k)$. Moreover, for any $a,b\in\bbR$ such that $\floor{a}\leq\ceil{b}$, we define $\intseq{a}{b}\eqdef\left\{\floor{a},\floor{a}+1,\dots,\ceil{b}-1,\ceil{b}\right\}$. Throughout the paper, $\log(\cdot)$ is \ac{wrt} base $e$, and therefore all the information quantities should be understood in \textit{nats}.

Let $\calD(\calH)$ denote the set of density operators acting on a \textit{separable} Hilbert space $\calH$. Let $\textnormal{id}_S$ denote the identity operator acting on $\calH_S$. The von Neumann entropy of $\hat{\rho}\in\calD(\calH)$ is $S(\hat{\rho})\eqdef-\tr{\hat{\rho}\log\hat{\rho}}$. The Holevo information of a \ac{c-q} ensemble $\left\{p_X(x),\hat{\rho}^x\right\}$ is $\chi\left(\left\{p_X(x),\hat{\rho}^x\right\}\right)\eqdef S\left(\sum_xp_X(x)\hat{\rho}^x\right)-\sum_xp_X(x)S(\hat{\rho}^x)$. In a bosonic system, self-adjoint canonical operators of position $\hat{q}$ and momentum $\hat{p}$ satisfy $[\hat{q},\hat{p}]=\imath\hbar$, where $\hbar$ is the reduced Planck constant and we let $\hbar=1$ in appropriate units. The creation and annihilation operators are defined as $\hat{a}\eqdef\frac{\hat{q}+\imath\hat{p}}{\sqrt{2}}$ and $\hat{a}^\dagger\eqdef\frac{\hat{q}-\imath\hat{p}}{\sqrt{2}}$, respectively. For $\alpha\in\bbC$ the one mode displacement operator is $\hat{D}(\alpha)\eqdef\exp\left(\alpha\hat{a}^\dagger-\alpha^*\hat{a}\right)$. The photon number operator is $\hat{N}\eqdef\hat{a}^\dagger\hat{a}$. Gordon's function is defined as $g(x)\eqdef(x+1)\log(x+1)-x\log x$. $\calL_{A\to BC}^{(\eta,N_B)}$ denotes the map corresponding to a bosonic channel with transmissivity $\eta$ and mixing thermal noise with mean photon number $N_B$. System $A$ is the input, system $B$ corresponds to the first output and system $C$ corresponds to the second output. $\calL^{(\eta,N_B)}_{A\to B}$ denotes the same bosonic channel as $\calL_{A\to BC}^{(\eta,N_B)}$ but the second output is traced out. The single-mode thermal state with mean photon number $N_\textnormal{th}$ is given by
\begin{align}
    \hat{\rho}_\textnormal{th}(N_\textnormal{th})
    &\eqdef
    \sum_{n=0}^\infty
    \frac{N_\textnormal{th}^n}{(N_\textnormal{th}+1)^{n+1}}
    \ket{n}\bra{n}\\
    &=
    \frac{1}{\pi N_\textnormal{th}}
    \int
    \exp\left(
        -\frac{\abs{\beta}^2}{N_\textnormal{th}}
    \right)
    \ket{\beta}\bra{\beta}\,\mathrm{d}^2\beta.
\end{align}
Quantum Chernoff bound is defined as $Q_s\left(\hat{\rho}\Vert\hat{\sigma}\right)\eqdef\tr{\hat{\rho}^s\hat{\sigma}^{1-s}}$ for $\hat{\rho},\hat{\sigma}\in\calD(\calH)$.

\subsection{Joint Communication and Sensing Model with Bipartite Entanglement Resource}\label{sec:model}
We consider the joint communication and sensing problem illustrated in~Figure 1, in which the physical system consists of two concatenated single-mode lossy thermal-noise bosonic channels, $\calL^{(\kappa,N_\textnormal{th}/(1-\kappa))}_{A\to CD}$ and $\calL^{(\tau,N_B)}_{C\to B}$. The first channel captures the effect of a target present in the link as a beamsplitter, where $\kappa$ denotes the transmissivity of the backscattered signal and $N_\textnormal{th}/(1-\kappa)$ is the mean photon number of the environmental thermal noise. The second channel models communication loss and noise between the transmitter and receiver, where $\tau$ is the transmissivity and $N_B$ is the mean photon number of the thermal noise. The transmitter (Alice) seeks to reliably communicate a message to a receiver (Bob) over multiple uses of the system while simultaneously estimating the range of the target through the backscattered signal, subject to an average mean photon number constraint $N$.

\begin{figure}[h]
    \centering
    \includegraphics[width=0.85\textwidth]{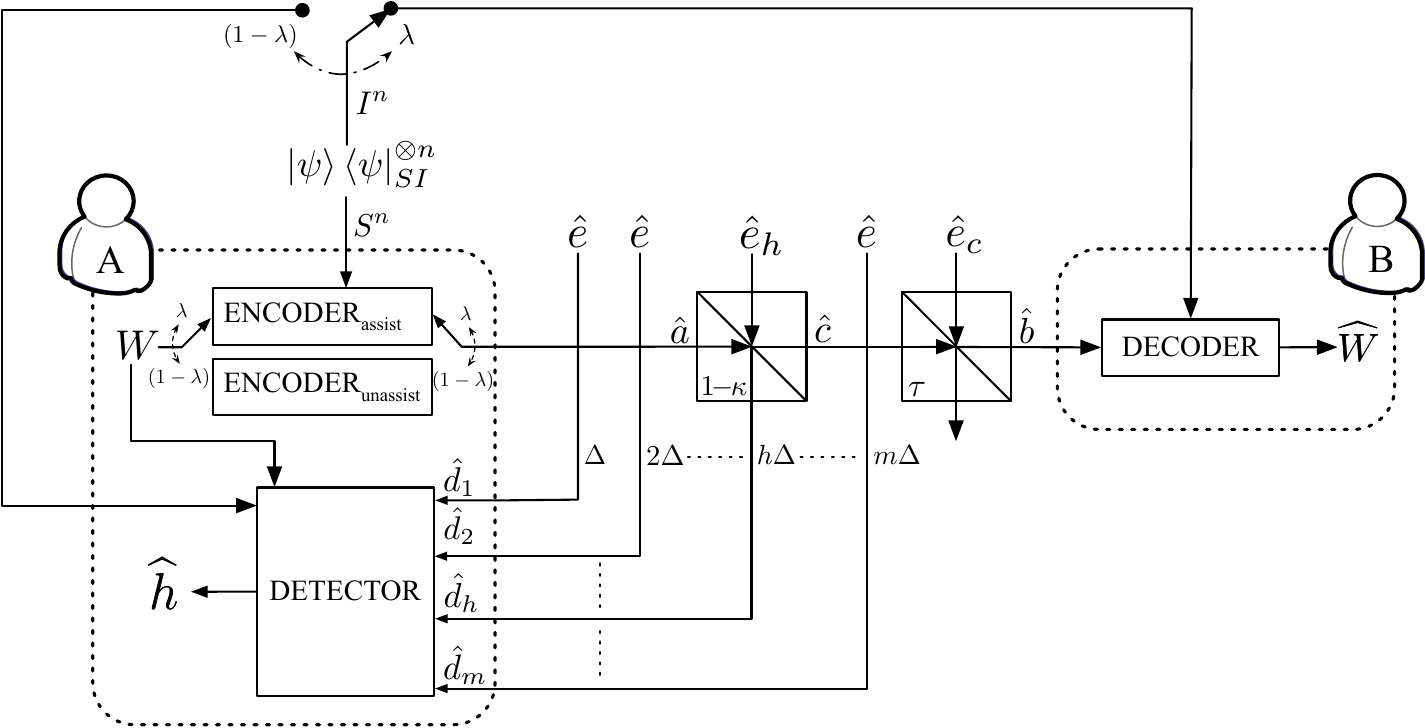}
    \caption{Joint communication and sensing model with bipartite entanglement resources over a lossy thermal-noise bosonic channel.}
    \label{fig:jcs_model}
\end{figure}
\FloatBarrier

Following the ranging formulation in~\cite{Zhuang2021}, the sensing task is modeled as an $m$-ary hypothesis testing problem. The known distance between Alice and Bob is partitioned into $m+1$ intervals of equal length $\Delta$, with $m\geq2$. Under hypothesis $h\in\intseq{1}{m}$, the beamsplitter is located at position $h\Delta$. The returned mode corresponding to the $i$-th interval arrives during the time window $t\in\left[2i\frac{\Delta}{c},\,2(i+1)\frac{\Delta}{c}\right)$, for $i\in\intseq{1}{m}$.

Under hypothesis $h$, the bosonic annihilation operators satisfy
\begin{align}
    \hat{d}_h &= -\sqrt{\kappa}\hat{a}+\sqrt{1-\kappa}\hat{e}_h,\\
    \hat{d}_i &= \hat{e}_i,\quad i\neq h,\\
    \hat{c} &= \sqrt{1-\kappa}\hat{a}+\sqrt{\kappa}\hat{e}_h,\\
    \hat{b} &= \sqrt{\tau}\hat{c}+\sqrt{1-\tau}\hat{e}_c.
\end{align}

As in~\cite{Zhuang2021,tan2008QI,Nair20}, we adopt the \ac{NPS} assumption. Under hypothesis $h$, the environmental mode $\hat{e}_h$ is a thermal state with mean photon number $N_\textnormal{th}/(1-\kappa)$, while each $\hat{e}_i$ for $i\neq h$ is a thermal state with mean photon number $N_\textnormal{th}$. The \ac{NPS} assumption corresponds to the indistinguishability of the hypotheses under vacuum input to the channel. Consequently, Alice must actively probe the channel in order to distinguish between the hypotheses. If the \ac{NPS} assumption is violated, then there is no quantum advantage in detection performance in the low-$N$ regime, since both the entanglement-assisted and unassisted error exponents converge to the constant vacuum-input performance as $N\to0$. The communication noise mode $\hat{e}_c$ is a thermal state with mean photon number $N_B$.

For notational simplicity, we define the effective communication transmissivity
\begin{align}
    \eta \eqdef \tau(1-\kappa),
\end{align}
and the effective communication noise photon number
\begin{align}
    N_T \eqdef \frac{\tau\kappa}{1-\kappa}N_\textnormal{th}+(1-\tau)N_B.
\end{align}

Alice has access to $n$ independent two-mode squeezed vacuum (TMSV) entangled state $\hat{\rho}_{SI}^{\otimes n}$, where $\hat{\rho}_{SI}=\ket{\psi}\bra{\psi}_{SI}$ and
\begin{align}
    \ket{\psi}_{SI}
    = \sum_{n=0}^\infty
    \sqrt{\frac{N_S^n}{(N_S+1)^{n+1}}}
    \ket{n}_S\ket{n}_I,
\end{align}
which she may either share with Bob to assist communication or retain to enhance sensing performance. We consider schemes that do not ``split'' the entanglement resource. The idler modes have to be partioned between Bob's decoder and Alice's detector. $\lambda\in[0,1]$ corresponds to the portion of the blocklength $n$ in which the idler modes are shared with Bob's decoder as described in Figure~\ref{fig:jcs_model}. Without loss of generality, we can assume consecutive modes.

Therefore, Alice's encoder is described by the encoder channels $\{\calE^{(w,\lambda)}_{S^n\to A^n}\otimes \textnormal{id}_{I^n}\}_{(w,\lambda)}$ mapping the message $w\in\intseq{1}{M}$ to a $2n$-mode state $\hat{\rho}_{A^nI^n}^{(w,\lambda)}$ entangled with the idler system $I^n$. Upon transmission through the concatenated channels
$\calL^{(\kappa,\,N_{\textnormal{th}}/(1-\kappa))}_{A\to CD}$
and
$\calL^{(\tau,\,N_B)}_{C\to B}$,
equivalently represented by the effective lossy thermal-noise channel $\calL^{(\eta,\,N_T)}_{A\to B}$, the resulting state is
\begin{align}
    \hat{\rho}_{B^n I^{1:\lambda n}}^{(w,\lambda)}
    \eqdef
    \tr[I^{\lambda n+1:n}]{
    \left(
    \left(\calL^{(\eta,\,N_T)}_{A\to B}\right)^{\otimes n}
    \otimes \textnormal{id}_{I^n}
    \right)
    \left(\hat{\rho}^{(w,\lambda)}_{A^n I^n}\right)}.
\end{align}
Bob then applies a \ac{POVM}
$\{\Pi^{(w,\lambda)}\}_{w=1}^M$
on the Hilbert space
$\calH_B^{\otimes n}\otimes\calH_I^{\otimes \lambda n}$
to decode the message $W$, producing an estimate $\widehat{W}$. Similarly, upon receiving the returned state under hypothesis $h\in\{1,\dots,m\}$, the joint state is
\begin{align}
    \hat{\rho}_{\{D_j^n\}_{j=1}^m\, I^{\lambda n+1:n}}^{(h,w,\lambda)}
    \eqdef\;
    &\left(
    \hat{\rho}_{\textnormal{th}}(N_{\textnormal{th}})^{\otimes n}
    \right)^{\otimes(h-1)}
    \otimes
    \tr[B^n,\,I^{1:\lambda n}]{\left(
    \left(\calL^{(\kappa,\,N_{\textnormal{th}}/(1-\kappa))}_{A\to BD}\right)^{\otimes n}
    \otimes \textnormal{id}_{I^n}
    \right)
    \left(\hat{\rho}^{(w,\lambda)}_{A^nI^n}\right)}
    \nonumber\\
    &\otimes
    \left(
    \hat{\rho}_{\textnormal{th}}(N_{\textnormal{th}})^{\otimes n}
    \right)^{\otimes(m-h)}.
\end{align}

Alice estimates the true hypothesis by applying the \ac{POVM}
$\{\Lambda^{(h,w,\lambda)}\}_{h\in\intseq{1}{m}}$
on the Hilbert space
$\left(\calH_D^{\otimes m}\otimes\calH_I\right)^{\otimes n}$,
producing an estimate $\widehat{h}$.

We define the communication rate as
\begin{align}
    R \eqdef \frac{\log M}{n},
\end{align}
and measure communication reliability using the maximum probability of error
\begin{align}
    P_{e,c}^n
    \eqdef
    \max_{w\in\intseq{1}{M}}
    \tr{
    \left(
    \textnormal{id}_{B^nI^{1:\lambda n}}
    -
    \Pi^{(w,\lambda)}
    \right)
    \hat{\rho}_{B^nI^{1:\lambda n}}^{(w,\lambda)}
    }.
\end{align}
We measure sensing performance using the maximum probability of error for hypothesis testing,
\begin{align}
    P_{e,d}^n
    \eqdef
    \max_{h\in\intseq{1}{m}}
    \max_{w\in\intseq{1}{M}}
    \tr{
    \left(
    \textnormal{id}_{\{D_j^n\}_{j=1}^mI^{\lambda n+1:n}}
    -
    \Lambda^{(h,w,\lambda)}
    \right)
    \hat{\rho}_{\{D_j^n\}_{j=1}^mI^{\lambda n+1:n}}^{(h,w,\lambda)}
    }.
\end{align}
The sensing error exponent is defined as
\begin{align}
    E
    \eqdef
    \lim_{n\to\infty}
    -\frac{1}{n}\log P_{e,d}^n.
\end{align}

As a necessary step toward establishing our main result, we require an extension of the multiple-hypothesis-testing Chernoff exponent developed in~\cite{Li2016Discriminating} to infinite-dimensional Hilbert spaces. Although one may project the density operators corresponding to each hypothesis onto finite-dimensional subspaces and establish convergence of the exponent through continuity properties of quantum $f$-divergences~\cite{Hiai2018}, one must still characterize the behavior of the missing mass induced by these projections. The analysis of this missing mass remains elusive. Nevertheless, the converse result of~\cite{Nussbaum2009Chernoff} extends to separable Hilbert spaces. Therefore, even though the achievability of the $M$-ary quantum Chernoff bound in infinite-dimensional settings remains conjecture, the converse still provides a fundamental performance limit. Accordingly, throughout this paper we assume that the multiple-hypothesis-testing Chernoff exponent is achievable for quantum states in infinite-dimensional Hilbert spaces, as has also been assumed in prior work~\cite{Zhuang2021}.

We restrict our attention to the low-signal-energy regime, in which $N\leq 1$, since this is the regime where the advantages provided by entanglement are most pronounced.

\section{Joint Communication and Sensing over Entanglement Assisted Bosonic Channels}
\label{sec:results}
We propose a joint communication and sensing protocol motivated by ring-state communication schemes for coherent states and \ac{TMSV} state~\cite{su2024}, and \ac{TMSV} assisted ranging~\cite{Zhuang2021}. Specifically, during a fraction $\lambda$ of the blocklength $n$, Alice shares the idler with Bob to assist communication, while during the remaining fraction $(1-\lambda)$ she retains the idler to improve ranging performance.

Alice's encoder consists of two parts:
\begin{enumerate}
    \item For the first $\lambda n$ modes, she encodes her classical message via \ac{PSK} by applying the single-mode phase-shift unitary
    \begin{align}
        \hat{U}_\theta \eqdef \exp\left(-2\imath\theta\hat{a}^\dagger\hat{a}\right)
    \end{align}
    to each signal mode.

    \item For the remaining $(1-\lambda)n$ modes, she encodes the message via displaced \ac{PSK} by applying the displacement unitary
    \begin{align}
        \hat{D}(\abs{\alpha}e^{\imath\theta})
    \end{align}
    to each signal mode.
\end{enumerate}
Here, $\theta\in[0,2\pi)$ is uniformly distributed in both schemes. Explicitly,
\begin{align}
    \calE^{(w,\lambda)}_{S^n\to A^n}
    =
    \left(\bigotimes_{j=1}^{\lambda n}\hat{U}_{\theta_j(w)}\right)
    \otimes
    \left(\bigotimes_{k=\lambda n+1}^{n}
    \hat{D}(\abs{\alpha}e^{\imath\theta_k(w)})\right).
\end{align}
We are implicitly assuming the existence of a classical local oscillator shared with Bob to be a phase reference for displaced \ac{PSK} scheme. The displacement modulation introduces the parameter
\begin{align}
    N_{\textnormal{m}}\eqdef \abs{\alpha}^2,
\end{align}
where ``m'' stands for modulation. The parameter $N_\textnormal{m}$ represents the energy injected into the channel by the displacement operation, and the total mean photon number constraint becomes
\begin{align}
    N_S+N_{\textnormal{m}}\leq N.
\end{align}

During the first $\lambda$ fraction of the blocklength, Alice knows the applied phase and therefore can eliminate it from the received state. The modes corresponding to all incorrect hypotheses are thermal states diagonal in the Fock basis and are therefore unaffected. During the remaning $(1-\lambda)$ fraction of the blocklength, Alice does not know which of the $\{\hat{d}_i\}_{i=1}^m$ is displaced so she cannot remove the displacement. However, she can still remove the phase of the displacement. After this phase cancellation, Alice's hypotheses become tensor products of two \ac{iid} backscattered states corresponding to the entanglement-unassisted and entanglement-assisted settings.

We first present a classical characterization of the achievable rate/error-exponent region for joint communication and sensing based coherent \ac{PSK}~\cite{su2024} and coherent detection~\cite{tan2008QI} in the following theorem, which serves as a baseline for the subsequent results.
\begin{theorem}
\label{thm:JBCS_class}
    The following rate/error-exponent region is achievable without entanglement assistance under a mean photon number constraint $N$:
    \begin{align}
        \calC_{\textnormal{classical}}= \left\{\begin{array}{l}
            (R,E)\in\bbR^2_+:\\
            R\leq R_\textnormal{classical}(N)\\
            E\leq E_\textnormal{classical}(N)
        \end{array}\right\},
    \end{align}
    where
    \begin{align}
        R_\textnormal{classical}(N) &= \frac{\eta N}{N_T+1}+\eta N\log\left(1+\frac{1}{N_T}\right)-(N_T+\eta N)\log\left(1+\frac{\eta N}{N_T(N_T+1)}\right),\\
        E_{\textnormal{classical}}(N) &= \frac{2\kappa N}{1+2N_\textnormal{th}+2\sqrt{N_\textnormal{th}(N_\textnormal{th}+1)}}.
    \end{align}
\end{theorem}
\begin{proof}
    See Section~\ref{sec:proof_class}.
\end{proof}

The set of achievable rate/error-exponent pairs in Theorem~\ref{thm:JBCS_class} forms a rectangular region. Consistent with classical information-theoretic results on joint communication and sensing over additive Gaussian channels~\cite{Joudeh2022}, there is no trade-off between the communication rate and the error exponent. In both~\cite{Joudeh2022} and the present model, communication is independent of the detection hypothesis.

Our main result on joint communication and sensing is the characterization of the achievable rate/error-exponent region with bipartite entanglement assistance, given in the following theorem.
\begin{theorem}
\label{thm:JBCS}
    Using the time-sharing between displaced-\ac{PSK} and \ac{TMSV}-\ac{PSK} schemes described earlier, the set of achievable rate/error-exponent pairs includes the set
    \begin{align}
        \calC_{\textnormal{JBCS}}=\bigcup_{\lambda,\delta\in[0,1]}\left\{\begin{array}{l}
            (R,E)\in\bbR^2_+:\\
            R\leq \lambda R_\textnormal{assist}(N)+(1-\lambda)R_\textnormal{unassist}(N,\delta)\\
            E\leq \lambda E_{\textnormal{unassist}}(N)+(1-\lambda)E_{\textnormal{assist}}(N,\delta)
        \end{array}\right\},
    \end{align}
    where
    \begin{align}
        R_\textnormal{assist}(N) &= g\left(N\right)+g\left(N_T\right)
	+\left(2 \eta N+N_T\right)N \log \left(\frac{\left(N_T+1\right) \left(N_T-\eta \right)}{N_T \left(N_T-\eta+1\right)}\right) \nonumber\\
&\quad
	+N \log \left(\frac{N_T+1}{N_T-\eta+1}\right)
	+\eta  N \log \left(\frac{\left(N_T+1\right)^2 \left(N_T-\eta \right)}{N_T^2 \left(N_T-\eta+1\right)}\right)\nonumber\\
	&\quad-g\left( \frac{\mu_+ - 1}{2}	\right)
	-g\left( \frac{\mu_- - 1}{2}  	\right),\\
    R_\textnormal{unassist}(N,\delta) &= \frac{\eta \delta N}{N_\textnormal{eff}+1}+\eta\delta N\log\left(1+\frac{1}{N_\textnormal{eff}}\right)\nonumber\\
    &-(\eta N+N_{T})\log\left(1+\frac{\eta\delta N}{N_\textnormal{eff}(N_\textnormal{eff}+1)}\right),\\
        E_{\textnormal{unassist}}(N) &= 2\log\left(\sqrt{(1+\kappa N+N_\textnormal{th})(1+N_{\textnormal{th}})}-\sqrt{\left(\kappa N+ N_\textnormal{th}\right)N_{\textnormal{th}}}\right),\\
        \hat{\rho}_{RI} &= \left(\calL^{(\kappa,N_\textnormal{th}/(1-\kappa))}_{A\to D}\otimes \textnormal{id}_{I}\right)\left(\hat{D}\left(\sqrt{2\delta N}\right)\otimes\textnormal{id}_I\hat{\rho}_{SI}D\left(\sqrt{2\delta N}\right)^\dagger\otimes\textnormal{id}_I\right),\\
        E_\textnormal{assist}(N,\delta) &= -\log Q_{1/2}\left(\hat{\rho}_{TRI}\Vert\hat{\rho}_{RTI}\right)\label{eq:QCB_TM_exp},
    \end{align}
    $N_\textnormal{eff}\eqdef \eta(1-\delta) N+N_T$, $\mu_{ \pm}\eqdef\sqrt{(N_T+(1+\eta)N+1)^2 - 4 \eta N (N+1)}\pm(N_T-(1-\eta) N)$, $\hat{\rho}_{TRI}=\hat{\rho}_{\textnormal{th}}\left(N_\textnormal{th}\right)\otimes\hat{\rho}_{RI}$ and $\hat{\rho}_{RTI}$ such that $\hat{\rho}_{RI}=\tr[T]{\hat{\rho}_{RTI}}$ and $\hat{\rho}_T=\tr[R,I]{\hat{\rho}_{RTI}}$.
\end{theorem}
\begin{proof}
    See Section~\ref{sec:proof_EA}.
\end{proof}

Note that the boundary of $\calC_\textnormal{JBCS}$ is attained when the mean photon number constraint is fully utilized, i.e., when $N_S + N_\textnormal{m} = N$.
Consequently, the constraint can be parameterized as $N_\textnormal{m} \eqdef \delta N$ and $N_S \eqdef (1 - \delta)N$ for $\delta \in [0,1]$. Moreover, when entanglement is reserved for ranging, the communication channel is equivalent to $\calL_{A\to B}^{\left(\eta, (\eta N_S+N_T)/(1-\eta)\right)}$. In this case, $N_S$ acts as additional noise, degrading the performance of unassisted communication. This can be further deduced by looking into $R_\textnormal{unassist}(N,\delta)$, in which $\eta N_S$ is acting as an effective thermal noise mean photon number and $N_\textnormal{m}$ acting as the channel input. Note also that $R_\textnormal{unassist}(N,1)=R_\textnormal{classical}(N)$ and $E_\textnormal{assist}(N,1)=E_\textnormal{classical}(N)$.

Turning to the error exponent, note that the expression~\eqref{eq:QCB_TM_exp} is computable using the techniques in~\cite{Pirandola2008}. One can deduce that the displacement term in the quantum Chernoff bound for Gaussian states~\cite{Pirandola2008} is quadratic in the differences of the mean vectors of the two hypotheses, which results in a linear dependence of the error exponent on $N_\textnormal{m}$. However, the mean-dependent part decreases with $N_S$, due to the inverse proportionality to the covariance matrices, while the mean independent part increases with $N_S$. This gives rise to a non-trivial trade-off between the communication rate and the sensing error-exponent \ac{wrt} to the energy split parameter $\delta$. 

Finally, we emphasize that $R_\textnormal{assist}(N)$ and $R_\textnormal{unassist}(N,\delta)$ are lower bounds on the Holevo information of the ensembles {\small $\left\{\frac{1}{2\pi},\left(\calL_{A\to B}^{(\eta,N_T)}\otimes\textnormal{id}_I\right)\left(\hat{U}_\theta\otimes\textnormal{id}_I\rho_{SI}\hat{U}_\theta^\dagger\otimes\textnormal{id}_I\right)\right\}_{\theta\in[0,2\pi)}$} and \par\noindent{\small $\left\{\frac{1}{2\pi},\calL_{A\to B}^{(\eta,N_T)}\left(\hat{D}(\sqrt{2\delta N}e^{i\theta})\rho_{S}\hat{D}(\sqrt{2\delta N}e^{i\theta})^\dagger\right)\right\}_{\theta\in[0,2\pi)}$}, respectively, obtained without invoking any Taylor series expansion in the physical parameters. Given the complexity of the joint communication and sensing problem described in Section~\ref{sec:model}, the expressions in Theorem~\ref{thm:JBCS} are remarkably compact.

Focusing on the region $N\ll1$, in which entanglement assistance provides the greatest enhancement, we first present how the assisted and unassisted communication rates and error-exponents are behaving in this regime in the following lemma.
\begin{lemma}\label{lm:low_N_rate_exp}
    In the regime $N\to0$, achievable rate and error-exponent terms $R_\textnormal{unassist}(N,\delta)$, $R_{\textnormal{assist}}(N)$, $E_{\textnormal{unassist}}(N)$ and $E_{\textnormal{assist},L}(N,\delta)$ behave as,
    \begin{align}
        R_{\textnormal{assist,L}}(N)
&= -\frac{\eta N \log N}{N_T+1}+N \left(
    (N_T+\eta)\log \frac{(N_T+1)(N_T-\eta)}{N_T(N_T-\eta+1)}\right. \nonumber\\
&\quad \left.+ \frac{\eta}{N_T+1} \left( 1+ \log \frac{(N_T+1)^2}{N_T(N_T-\eta+1)}\right)
\right),\\
        R_{\textnormal{unassist,L}}(N,\delta) &= \delta\eta N\left(
        \log\left(1+\frac{1}{N_T+1}\right)-\frac{1}{N_T(N_T+1)}\right),\\
        E_\textnormal{unassist,L}(N) &= \frac{\kappa^2N^2}{4 N_\textnormal{th}(N_\textnormal{th}+1)},\\
        E_\textnormal{assist,L}(N,\delta)&=2\kappa N\left(\frac{1-\delta}{N_{\textnormal{th}}+1}+\frac{\delta}{1+2N_\textnormal{th}+2\sqrt{N_\textnormal{th}(N_\textnormal{th}+1)}}\right).
    \end{align}
\end{lemma}
\begin{proof}
    See Section~\ref{sec:proof_EA}.
\end{proof}

A simplified and more explicit achievable rate/error-exponent region is given in the following theorem.

\begin{corollary}
\label{cor:JBCS_approx}
    Using the time-sharing between displaced-\ac{PSK} and \ac{TMSV}-\ac{PSK} schemes, in the regime $N\ll1$, the set of achievable rate/error-exponent pairs includes the set
    \begin{align}
        \calC_{\textnormal{JBCS}} = \bigcup_{\lambda\in[0,1]}\left\{\begin{array}{l}
            (R,E)\in\bbR^2_+:\\
            R\leq \lambda R_\textnormal{assist,L}(N)\\
            E\leq \lambda E_{\textnormal{unassist,L}}(N)+(1-\lambda)E_{\textnormal{assist,L}}(N)
        \end{array}\right\},
    \end{align}
    where
    \begin{align}
R_{\textnormal{assist,L}}(N)
&= -\frac{\eta N \log N}{N_T+1}+N \left(
    (N_T+\eta)\log \frac{(N_T+1)(N_T-\eta)}{N_T(N_T-\eta+1)}\right. \nonumber\\
&\quad \left.+ \frac{\eta}{N_T+1} \left( 1+ \log \frac{(N_T+1)^2}{N_T(N_T-\eta+1)}\right)
\right),\\
        E_\textnormal{unassist,L}(N) &= \frac{\kappa^2N^2}{4 N_\textnormal{th}(N_\textnormal{th}+1)},\\
        E_\textnormal{assist,L}(N)&=\frac{2\kappa N}{N_{\textnormal{th}}+1}.
    \end{align}
\end{corollary}
\begin{proof}
    See Section~\ref{sec:proof_EA}.
\end{proof}

Note that $\calC_\textnormal{JBCS}$ reduces to time-sharing between the rate/error-exponent pairs $(R_\textnormal{assist,L},E_{\textnormal{unassist,L}})$ and $(0,E_{\textnormal{assist,L}})$, which enforces $\delta=0$ in Theorem~\ref{thm:JBCS}. This follows from the imbalance between the scalings of $R_\textnormal{assist}(N)=\calO\left(-N\log N\right)$ and $R_\textnormal{unassist}(N,\delta)=\calO\left(N\right)$ established in Lemma~\ref{lm:low_N_rate_exp}. As a result, the time-sharing curve has slope on the order of $\calO\left(-1/\log N\right)$, whereas the curve $\{E_\textnormal{assist}(N,\delta)\}_{\delta\in[0,1]}$ has slope on the order of $\calO(1)$. Since both slopes are negative, the $(N_S,N_\textnormal{m})$-split curve always lies below the time-sharing curve. Also, $E_\textnormal{assist,L}(N)$ is consistent with the ranging exponent in~\cite{Zhuang2021} as $N_\textnormal{th}\to\infty$.

\begin{figure}[h]
    \centering
    \includegraphics[width=\linewidth]{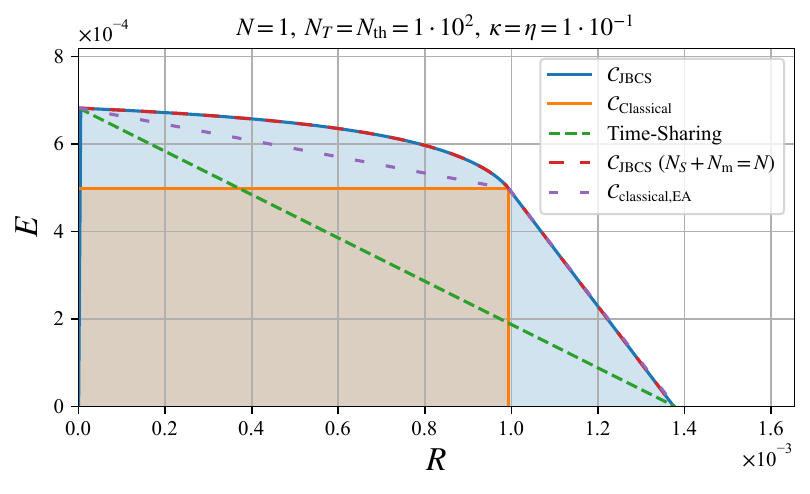}
    \caption{Comparison of the regions corresponding to Theorems~\ref{thm:JBCS} and \ref{thm:JBCS_class} against Time-Sharing.}
    \label{fig:JBCS_tradeoff}
\end{figure}

We illustrate Theorems~\ref{thm:JBCS_class} and \ref{thm:JBCS} through several numerical examples. Figure~\ref{fig:JBCS_tradeoff} presents the achievable joint communication and sensing rate/error-exponent regions comparing $\calC_{\textnormal{classical}}$, $\calC_{\textnormal{JBCS}}$, and the exact time-sharing scheme between the \ac{QI} exponent~\cite{Zhuang2021}, denoted by
\begin{align}
    E_{\textnormal{QI}}(N)\eqdef E_{\textnormal{assist}}(N,0),
\end{align}
attained at zero rate $(R,E)=(0,E_{\textnormal{QI}})$, and the entanglement-assisted capacity~\cite{Bennett2002}
\begin{align}
    C_{\textnormal{EA}}(N)
    = g(N)+g(\eta N+N_T)
    -g\left(\frac{\mu_{+}-1}{2}\right)
    -g\left(\frac{\mu_{-}-1}{2}\right),
\end{align}
attained at the unassisted exponent $E_{\textnormal{unassist}}(N)$.

Under the mean photon number constraint $N=1$, Figure~\ref{fig:JBCS_tradeoff} illustrates the comparison between the achievable joint communication and sensing regions. Note that $\calC_\textnormal{JBCS}$ is characterized by taking the union over all $N_S+N_\textnormal{m}\leq N, N_S,N_\textnormal{m}\in[0,N]$. The red curve characterizing the boundary of $\calC_{\textnormal{JBCS}}$ is attained by fully utilizing the mean photon number constraint, i.e.,
\begin{align}
    N_S+N_\textnormal{m}=N.
\end{align}
Both $\calC_{\textnormal{classical}}$ and $\calC_{\textnormal{JBCS}}$ outperform exact time-sharing. Moreover, $\calC_{\textnormal{JBCS}}$ contains the convex hull of
\begin{align}
    \calC_{\textnormal{classical}}
    \cup
    \left\{
        (0,0),(0,E_{\textnormal{QI}}(N)),
        (C_{\textnormal{EA}}(N),E_{\textnormal{unassist}}(N)),
        (C_{\textnormal{EA}}(N),0)
    \right\},
\end{align}
which is denoted in Figure~\ref{fig:JBCS_tradeoff} as $\calC_{\textnormal{JBCS,EA}}$, demonstrating a clear quantum advantage over classical performance.

\begin{figure}[h]
    \centering
    \includegraphics[width=\linewidth]{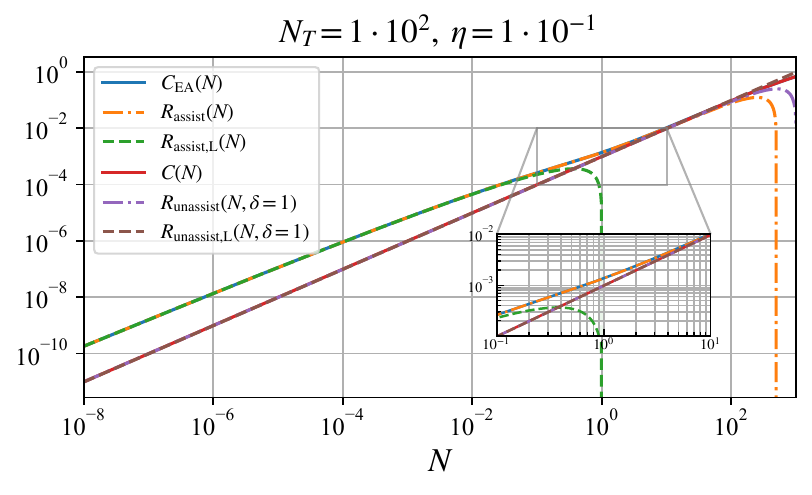}
    \caption{Comparison of rates in Theorems~\ref{thm:JBCS_class} and \ref{thm:JBCS}, their approximations in Lemma~\ref{lm:low_N_rate_exp} against $C_{\textnormal{EA}}$ and $C$.}
    \label{fig:rate_comparison}
\end{figure}

\begin{figure}[h]
    \centering
    \includegraphics[width=\linewidth]{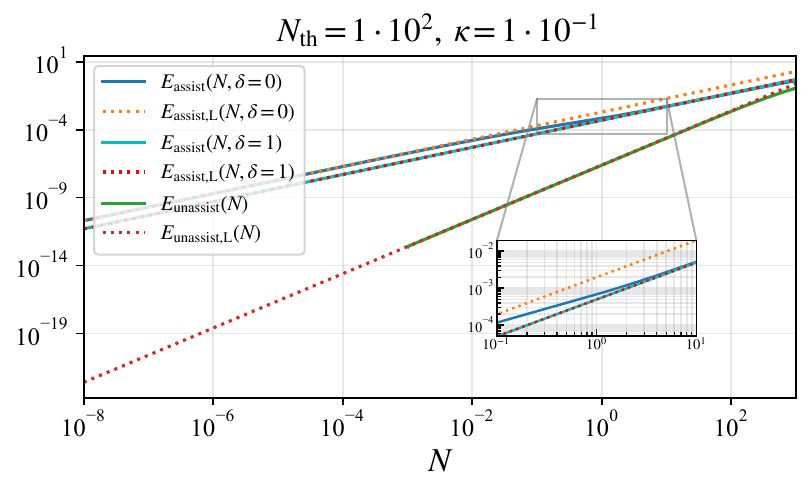}
    \caption{Comparison of exponents in Theorems~\ref{thm:JBCS_class} and \ref{thm:JBCS}, their approximations in Lemma~\ref{lm:low_N_rate_exp}.}
    \label{fig:exponents}
\end{figure}

The achievable rate/error-exponent regions shown in Figure~\ref{fig:JBCS_tradeoff} correspond to the zoomed-in windows presented in Figures~\ref{fig:rate_comparison} and~\ref{fig:exponents}. In this regime, $R_\textnormal{unassist}(N,1)$ closely approximates the classical capacity
\begin{align}
    C(N)=g(\eta N+N_T)-g(N_T),
\end{align}
while $R_{\textnormal{assist}}(N)$ closely approximates $C_{\textnormal{EA}}(N)$. At the same time, although $C_{\textnormal{EA}}(N)$ and $C(N)$ converge toward each other, a non-negligible quantum advantage remains. Similarly, the exponents $E_{\textnormal{assist}}(N,0)$ and $E_{\textnormal{assist}}(N,1)$ also converge, while still maintaining a quantum advantage. This convergence between assisted and unassisted rates/error-exponents induces a nontrivial trade-off between $N_S$ and $N_\textnormal{m}$, leading to the advantage of $\calC_{\textnormal{JBCS}}$ over both $\calC_\textnormal{classical}$ and exact time-sharing.

\begin{figure}[h]
    \centering
    \includegraphics[width=\linewidth]{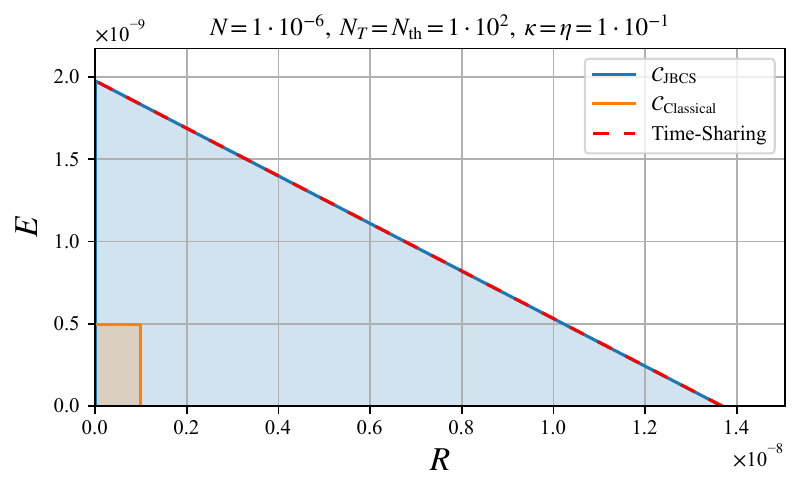}
    \caption{Comparison of the regions corresponding to Theorem~\ref{thm:JBCS_class} and Corollary~\ref{cor:JBCS_approx} against Time-Sharing.}
    \label{fig:JBCS_no_tradeoff}
\end{figure}

Figure~\ref{fig:JBCS_no_tradeoff} presents the achievable joint communication and sensing regions in the low mean photon number regime $N=10^{-6}$, comparing $\calC_{\textnormal{classical}}$, $\calC_{\textnormal{JBCS}}$, and exact time-sharing. In this regime, $\calC_\textnormal{JBCS}$ collapses to the time-sharing region, while $\calC_\textnormal{classical}$ becomes a small rectangular region contained inside it, illustrating a strong quantum advantage over classical performance. Furthermore, Figures~\ref{fig:rate_comparison} and~\ref{fig:exponents} indicate that the leading-order approximations in Lemma~\ref{lm:low_N_rate_exp} are tight in this regime.

\FloatBarrier

\FloatBarrier

\FloatBarrier

\FloatBarrier

\section{Proofs}
\label{sec:proofs}
We divide the proofs of Theorem~\ref{thm:JBCS}, Lemma~\ref{lm:low_N_rate_exp} and Corollary~\ref{cor:JBCS_approx} into analyses of the individual communication and sensing tasks when entanglement is shared for communication or kept for sensing. Specifically, we first analyze the cases in which entanglement is shared exclusively with either the transmitter or the receiver, and then show how these results can be combined to obtain the achievable joint communication and sensing region. We begin with the proof of Theorem~\ref{thm:JBCS_class}.

\subsection{Proof of Theorem~\ref{thm:JBCS_class}}\label{sec:proof_class}

\begin{itemize}
    \item \textit{Classical Communication:} We consider Alice using the ensemble
    $\left\{\frac{1}{2\pi},\ket{\sqrt{N}e^{\imath\theta}}\bra{\sqrt{N}e^{\imath\theta}}\right\}_{\theta\in[0,2\pi)}$
    for classical communication. Let
    \[
    \hat{\rho}_B^\theta
    \eqdef
    \calL_{A\to B}^{(\eta,N_T)}
    \left(
    \ket{\sqrt{N}e^{\imath\theta}}
    \bra{\sqrt{N}e^{\imath\theta}}
    \right).
    \]
    We use the following lemma from~\cite{su2024} as a lower bound on the achievable communication rate.

\begin{lemma}[\cite{su2024}]\label{lm:coherent_PSK}
    The Holevo information of the ensemble
    $\left\{\frac{1}{2\pi},\rho_B^\theta\right\}_{\theta\in[0,2\pi)}$
    is lower bounded by
    \begin{align}
        \chi\left(\left\{\frac{1}{2\pi},\hat{\rho}_B^\theta\right\}_{\theta\in[0,2\pi)}\right)
        &\geq
        \frac{\eta N}{N_T+1}
        +\eta N\log\left(1+\frac{1}{N_T}\right)\nonumber\\
        &\quad
        -(N_T+\eta N)\log\left(1+\frac{\eta N}{N_T(N_T+1)}\right),
    \end{align}
    for all $N,N_{\textnormal{th}}\in\bbR_+$ and $\eta\in[0,1]$.
\end{lemma}

Note that Lemma~\ref{lm:coherent_PSK} holds without requiring approximations or assumptions such as the low-signal or high-noise regime.

    \item \textit{Classical Sensing:} Let
    $\hat{\rho}_\textnormal{th}$ denote
    $\hat{\rho}_\textnormal{th}\left(N_\textnormal{th}\right)$,
    and let $\hat{D}_\alpha$ denote $\hat{D}(\alpha)$.
    Since Alice knows the phase of the displacement, she can eliminate the phase in the returned signal at the detector. This affects only the true hypothesis, since it is the only state with non-zero mean, whereas the remaining states are thermal states diagonal in the Fock basis.

    The classical sensing problem is a hypothesis test between the states
    \begin{align}
    \left\{\hat{\rho}_h^{\otimes n}\right\}_{h=1}^m
    \eqdef
    \left\{
    \left(\hat{\rho}_\textnormal{th}^{\otimes n}\right)^{\otimes (h-1)}
    \otimes
    \hat{\rho}_D^{\otimes n}
    \otimes
    \left(\hat{\rho}_\textnormal{th}^{\otimes n}\right)^{\otimes (m-h)}
    \right\}_{h=1}^{m},
    \end{align}
    where
    \begin{align}
    \hat{\rho}_D
    &\eqdef
    \calL^{(\kappa,N_\textnormal{th}/(1-\kappa))}
    \left(
    \ket{\sqrt{N}}\bra{\sqrt{N}}
    \right)\\
    &=
    \hat{D}_{\sqrt{\kappa N}}
    \hat{\rho}_\textnormal{th}
    \hat{D}_{\sqrt{\kappa N}}^\dagger\label{eq:rho_D}.
    \end{align}

    From the $M$-ary Chernoff bound~\cite{Li2016Discriminating} and defining the unitary $\hat{U}_{\sqrt{\kappa N}}\eqdef \hat{D}_{\sqrt{\kappa N}}\otimes \hat{D}_{\sqrt{\kappa N}}^\dagger$,
    \begin{align}
        E_\textnormal{classical}
        &=
        \min_{(i,j):i\neq j}
        \max_{s\in[0,1]}
        -\log Q_s\left(\hat{\rho}_i\Vert\hat{\rho}_j\right)\\
        &\overset{(a)}{=}
        \max_{s\in[0,1]}
        -\log
        Q_s\left(
        \hat{\rho}_\textnormal{th}\otimes\hat{\rho}_D
        \Vert
        \hat{\rho}_D\otimes\hat{\rho}_\textnormal{th}
        \right)\\
        &\overset{(b)}{=}
        \max_{s\in[0,1]}
        -\log
        Q_s\left(\hat{U}_{\sqrt{\kappa N}}
        \left(\hat{\rho}_D\otimes\hat{\rho}_\textnormal{th}\right)\hat{U}_{\sqrt{\kappa N}}^\dagger
        \big\Vert
        \hat{\rho}_D\otimes\hat{\rho}_\textnormal{th}\right)\\
        &\overset{(c)}{=}
        \max_{s\in[0,1]}
        -\log
        Q_{s}\left(\hat{\rho}_D\otimes\hat{\rho}_\textnormal{th}
        \big\Vert
        \hat{U}_{\sqrt{\kappa N}}^\dagger\left(\hat{\rho}_D\otimes\hat{\rho}_\textnormal{th}
        \right)\hat{U}_{\sqrt{\kappa N}}\right)\\
        &=
        \max_{s\in[0,1]}
        -\log
        Q_{1-s}\left(\hat{\rho}_\textnormal{th}\otimes\hat{\rho}_D
        \Vert
        \hat{\rho}_D\otimes\hat{\rho}_\textnormal{th}
        \right)\\
        &=
        -2\log
        Q_{1/2}
        \left(
        \hat{\rho}_D
        \Vert
        \hat{\rho}_\textnormal{th}
        \right).
        \label{eq:e_classic}
    \end{align}

    Here, (a) follows from the symmetry of the problem, similarly to~\cite[Eq.~(5)]{Zhuang2021}, (b) follows from $\hat{\rho}_D$ in~\eqref{eq:rho_D} and definition of the unitary $\hat{U}_{\sqrt{\kappa N}}$, and (c) follows from the unitary invariance of $Q_s$.
    By uniqueness of the optimizing $s$~\cite{Audenaert2007}, we obtain $s^*=1/2$.
    Using techniques from~\cite{Pirandola2008}, \eqref{eq:e_classic} simplifies to
    \begin{align}
        E_\textnormal{classical}(N)
        =
        \frac{2\kappa N}
        {1+2N_\textnormal{th}+2\sqrt{N_\textnormal{th}(N_\textnormal{th}+1)}},
    \end{align}
    which is consistent with the optimal classical ranging exponent in~\cite[Eq.~(5)]{Zhuang2021} for general $N_\textnormal{th}\in\bbR_+$.
\end{itemize}

\subsection{Proofs of Theorem~\ref{thm:JBCS}, Lemma~\ref{lm:low_N_rate_exp} and Corollary~\ref{cor:JBCS_approx} }\label{sec:proof_EA}
\begin{itemize}
    \item \textit{Unassisted Communication:} For unassisted communication, the encoder applies displaced phase modulation to the signal mode of the \ac{TMSV} state. The displacement is crucial because, from Bob's perspective, the signal state is a thermal state with mean photon number $N_S$, and pure phase modulation alone does not modify the state. Formally, the displaced phase modulation is defined as
\begin{align}
    \hat{\rho}_{A}^\theta &= \hat{D}\left(\sqrt{N}_\textnormal{m}e^{\imath\theta}\right)\hat{\rho}_{S}\hat{D}\left(\sqrt{N}_\textnormal{m}e^{\imath\theta}\right)^\dagger,
\end{align}
where $\hat{\rho}_{S}=\tr[I]{\ket{\psi}\bra{\psi}_{SI}}$ and $\theta\in[0,2\pi)$. Bob's channel output is
\begin{align}
    \hat{\rho}_B^\theta &= \hat{D}(\sqrt{\eta}\alpha_\theta)\hat{\rho}_{\textnormal{th}}(N_\eta)\hat{D}(\sqrt{\eta}\alpha_\theta)^\dagger.
\end{align}
Using Lemma~\ref{lm:coherent_PSK} with $N_T$ replaced by $N_\eta=\eta N_S+N_T$, the Holevo information of the displaced \ac{PSK} scheme is lower bounded by
\begin{align}
    \chi\left(\left\{\frac{1}{2\pi},\hat{\rho}_B^\theta\right\}\right)&\geq\frac{\eta N_\textnormal{m}}{N_\eta+1}+\eta N_\textnormal{m}\log\left(1+\frac{1}{N_\eta}\right)\nonumber\\
        &-(N_\eta+\eta N_\textnormal{m})\log\left(1+\frac{\eta N_\textnormal{m}}{N_\eta(N_\eta+1)}\right)\label{eq:rate_dpsk},
\end{align}
$\forall N_S,N_\textnormal{m},N_{\textnormal{th}}\in\bbR_+$ and $\eta\in[0,1]$.

In the regime $N\ll 1$, substituting the parameterization of the boundary $N_S=(1-\delta)N,N_\textnormal{m}=\delta N$ and taking the Taylor expansion around $N=0$ we obtain the following approximation as the leading order term
\begin{align}
    R_\textnormal{unassist,L}(N,\delta) = \delta\eta N\left(
        \log\left(1+\frac{1}{N_T+1}\right)-\frac{1}{N_T(N_T+1)}\right)+\calO\left(N^2\right)\label{eq:r_unassist_l}.
\end{align}

    \item \textit{Assisted Sensing:} As described in the time-sharing protocol between displaced-\ac{PSK} and \ac{TMSV}-\ac{PSK} protocol in Section~\ref{sec:results}, Alice applies a phase modulation to the received modes in the detector to eliminate the phase. Hence the mean and covariance of the true hypothesis become
\begin{align}
    r_{RI} &= -\sqrt{2\kappa N_{\textnormal{m}}}\begin{bmatrix}
        1\\
        0\\
        0\\
        0
    \end{bmatrix}\\
    V_{RI} &= \begin{bmatrix}
        E\bfI & C\bfZ\\
        C\bfZ & S\bfI
    \end{bmatrix}
\end{align}
where $\bfZ=\begin{bmatrix}1 & 0\\0 & -1\end{bmatrix},E=2(\kappa N_S+N_\textnormal{th})+1$, $C=-2\sqrt{\kappa N_S(N_S+1)}$ and $S=2N_S+1$.

The detection problem therefore reduces to hypothesis testing between the states
\begin{align}
    \{\hat{\rho}_h^{\otimes n}\}_{h=1}^m\eqdef\left\{\left(\hat{\rho}_{\textnormal{th}}^{\otimes (h-1)}\right)^{\otimes n}\otimes\hat{\rho}_{RI}^{\otimes n}\otimes\left(\hat{\rho}_{\textnormal{th}}^{\otimes (m-h)}\right)^{\otimes n}\right\}_{h=1}^m.
\end{align}

By applying $M$-ary Quantum Chernoff bound, the error exponent is given as
\begin{align}
    E_\textnormal{assist}(N_S,N_{\textnormal{m}}) &= \min_{(i,j):i\neq j}\max_{s\in[0,1]}-\log Q_s\left(\hat{\rho}_i\Vert\hat{\rho}_j\right)\\
    &=\max_{s\in[0,1]}-\log Q_s\left(\hat{\rho}_{\textnormal{th}}(N_{\textnormal{th}})\otimes\hat{\rho}_{RI}\Vert \hat{\rho}_{RI}\otimes \hat{\rho}_{\textnormal{th}}(N_{\textnormal{th}})\right)\\
    &=\max_{s\in[0,1]}-\log Q_s\left(\hat{\rho}_{TRI}\Vert \hat{\rho}_{RTI}\right).
\end{align}
The corresponding covariance matrices and means of the states are
\begin{align}
    V_{TRI} &= \begin{bmatrix}
        (2N_\textnormal{th}+1)\bfI & \bfzero & \bfzero\\
        \bfzero & E\bfI & C\bfZ\\
        \bfzero & C\bfZ & S\bfI
    \end{bmatrix},\quad r_{TRI} = -\sqrt{2\kappa N_\textnormal{m}}\begin{bmatrix}
        1\\0\\0\\0\\0\\0
    \end{bmatrix},\\
    V_{RTI} &= \begin{bmatrix}
        E\bfI & \bfzero & C\bfZ\\
        \bfzero & (2N_\textnormal{th}+1)\bfI & \bfzero\\
        C\bfZ & \bfzero & S\bfI
    \end{bmatrix},\quad r_{RTI} = -\sqrt{2\kappa N_\textnormal{m}}\begin{bmatrix}
        0\\0\\1\\0\\0\\0
    \end{bmatrix}.
\end{align}
Notice that, for $P=\begin{bmatrix}
    \bfzero &\bfI&\bfzero\\
    \bfI&\bfzero &\bfzero\\
    \bfzero&\bfzero&\bfI
\end{bmatrix}$ we have $V_{RTI} = PV_{TRI}P^T$. Since $P$ is a symplectic matrix there exists a Gaussian unitary $\hat{U}_{RT}$ such that $\hat{\rho}_{RTI}=\hat{U}_{RT}\hat{\rho}_{TRI}\hat{U}_{RT}^\dagger$. Similar to the classical exponent
\begin{align}
    \max_{s\in[0,1]}-\log Q_s\left(\hat{\rho}_{TRI}\Vert\hat{\rho}_{RTI}\right) &= \max_{s\in[0,1]}-\log\tr{\hat{\rho}_{TRI}^s\left(\hat{U}_{RT}\hat{\rho}_{TRI}\hat{U}_{RT}^\dagger\right)^{1-s}}\\
    &= \max_{s\in[0,1]}-\log\tr{\left(\hat{U}_{RT}^\dagger\hat{\rho}_{TRI}\hat{U}_{RT}\right)^s\hat{\rho}_{TRI}^{1-s}}\\
    &= \max_{s\in[0,1]}-\log Q_{1-s}\left(\hat{\rho}_{TRI}\Vert\hat{\rho}_{RTI}\right).
\end{align}
Once again, by symmetry of the problem, the optimal value is $s=1/2$.

Following the approach of~\cite{Pirandola2008}, the Chernoff bound between the two Gaussian states with covariance matrices $(V_{RTI},r_{RTI})$ and $(V_{TRI},r_{TRI})$, we need the symplectic decompositions $(S_1,\{\nu_\textnormal{th},\nu_1,\nu_2\})$ and $(PS_1P^T,\{\nu_1,\nu_\textnormal{th},\nu_2\})$, respectively, in which 
\begin{align}
    \nu_i=((-1)^{i+1}(N_{\textnormal{th}}-(1-\kappa)N_S)+\sqrt{(1+(1+\kappa)N_S+N_{\textnormal{th}})^2-4\kappa N_S(N_S+1)}).
\end{align}
for $i\in\{1,2\}$~\cite{tan2008QI} and $\nu_\textnormal{th}=2N_\textnormal{th}+1$. The symplectic matrix $S_1$ that diagonalizes $V_{TRI}$ is given as
\begin{align}
    S_1 &= \begin{bmatrix}
        \bfI &\bfzero&\bfzero\\
        \bfzero& x_+\bfI& x_-\bfZ\\
        \bfzero& x_+\bfZ& x_+\bfI
    \end{bmatrix},\\
    x_\pm &=\sqrt{\frac{1+(1+\kappa)N_S+N_{\textnormal{th}}\pm\sqrt{(1+(1+\kappa)N_S+N_{\textnormal{th}})^2-4\kappa N_S(N_S+1)}}{2\sqrt{(1+(1+\kappa)N_S+N_{\textnormal{th}})^2-4\kappa N_S(N_S+1)}}}.
\end{align}
Then the Quantum Chernoff bound~\cite{Pirandola2008} is given as
\begin{align}
    Q_{1/2} &= \overline{Q}_{1/2}\exp\left(-d^T(V_1+V_2)^{-1}d\right),\label{eq:QCB}
\end{align}
where
\begin{align}
    V_1 &= S_1\left( \Lambda\left(\nu_\textnormal{th}\right)\bfI_2\oplus\Lambda\left(\nu_1\right)\bfI_2\oplus\Lambda\left(\nu_2\right)\bfI_2\right)S_1^T,\\
    d &= r_{TRI}-r_{RTI}\nonumber\\
      &= \sqrt{2\kappa N_\textnormal{m}}\begin{bmatrix}
        1 & 0 & -1 & 0 & 0 & 0
    \end{bmatrix}^T\label{eq:mean_diff},\\
    V_2 &= PV_1P^T,\\
    \overline{Q}_{1/2} &= \frac{\Lambda\left(\nu_1\right)\Lambda\left(\nu_2\right)\Lambda\left(\nu_\textnormal{th}\right)}{2E_R(E_RS_R-C_R^2)},\\
    E_R&\eqdef\Lambda\left(\nu_\textnormal{th}\right)+x_+^2\Lambda\left(\nu_1\right)+x_-^2\Lambda\left(\nu_2\right),\\
    S_R &\eqdef x_-^2\Lambda\left(\nu_1\right)+x_+^2\left(\nu_2\right),\\
    C_R &\eqdef x_+x_-\left(\Lambda\left(\nu_1\right)+\Lambda\left(\nu_2\right)\right),\\
    \det{V_1+V_2} &= \left(2E_R\left(E_RS_R-C_R^2\right)\right)^2.
\end{align}
According to the parameterization $N_\textnormal{m}=\delta N$, $N_S=(1-\delta)N$, the Taylor expansion of the error-exponent corresponding to mean independent part of quantum Chernoff bound in~\cite{Pirandola2008}, denoted $-\log \overline{Q}_{1/2}$, around $N=0$ yields
\begin{align}
    -\log\overline{Q}_\frac{1}{2}&=\frac{2\kappa(1-\delta) N}{1+N_{\textnormal{th}}}+o\left(N\right).
\end{align}
Taylor series of the displacement term around $N=0$ is given by
\begin{align}
    \frac{1}{2}d^T\left(V_1+V_2\right)^{-1}d &= \frac{2\delta\kappa N}{1+2N_{\textnormal{th}}+2\sqrt{N_{\textnormal{th}}(N_{\textnormal{th}}+1)}}+\calO\left(N^2\right).
\end{align}

Hence, the error exponent for $N\ll1$ becomes 
\begin{align}
    E_\textnormal{assist}(N,\delta) &= 2\kappa N\left(\frac{\delta}{{N_{\textnormal{th}}}+1}+\frac{1-\delta}{1+2N_{\textnormal{th}}+2\sqrt{N_{\textnormal{th}}(N_{\textnormal{th}}+1)}}\right)+\calO\left(N^2\right)\label{eq:exp_assist_l}.
\end{align}

    \item \textit{Assisted Communication:} In order the lower bound the achievable rate using phase modulated TMSV we use the following lemma from~\cite{su2024}, in which $\hat{\rho}_{BI}^\theta=\hat{U}_\theta\otimes \textnormal{id}_I\left(\calL^{(\eta,N_T/(1-\eta))}\otimes \textnormal{id}_I\right)\left(\rho_{SI}\right)\hat{U}_\theta^\dagger\otimes \textnormal{id}_I$
\begin{lemma}[\cite{su2024}]
    Under uniform phase modulation of \ac{TMSV}, the Holevo information of the ensemble $\left\{\frac{1}{2\pi},\hat{\rho}_{BI}^\theta\right\}_{\theta\in[0,2\pi)}$ is lower bounded by
    \begin{align}
        \chi\left(\left\{\frac{1}{2\pi},\hat{\rho}_{BI}^\theta\right\}_{\theta\in[0,2\pi)}\right)&\geq g\left(N\right)+g\left(N_T\right)
	+\left(2 \eta N+N_T\right)N \log \left(\frac{\left(N_T+1\right) \left(N_T-\eta \right)}{N_T \left(N_T-\eta+1\right)}\right) \\
&\quad
	+N \log \left(\frac{N_T+1}{N_T-\eta+1}\right)
	+\eta  N \log \left(\frac{\left(N_T+1\right)^2 \left(N_T-\eta \right)}{N_T^2 \left(N_T-\eta+1\right)}\right)\\
	&-g\left( \frac{\mu_+ - 1}{2}	\right)
	-g\left( \frac{\mu_- - 1}{2}  	\right),
    \end{align}
    $\forall N,N_T\in\bbR_+$ such that $N_T>\max(\eta,\eta N-1)$ and $\eta\in[0,1]$ where $\mu_\pm = \pm(N_T-(1-\eta)N)+\sqrt{(1+(1+\eta)N+N_T)-4\eta N(N+1)}$.
    \label{lm:HI_bound_assisted}
\end{lemma}

In the regime $N\ll 1$, using the lemma from~\cite{su2024},
\begin{lemma}[\cite{su2024}]
    Taylor series of the lower bound of Holevo information around $N=0$ is
    \begin{align}
        -\frac{\eta N \log N}{N_T+1}+N \left(
    (N_T+\eta)\log \frac{(N_T+1)(N_T-\eta)}{N_T(N_T-\eta+1)}+ \frac{\eta}{N_T+1} \left( 1+ \log \frac{(N_T+1)^2}{N_T(N_T-\eta+1)}\right)\label{eq:r_assist_L}
\right).
    \end{align}
\end{lemma}
We denote \eqref{eq:r_assist_L} by $R_\textnormal{assist,L}(N)$.
    \item \textit{Unassisted Sensing:} Since we are using phase modulation for entanglement assisted communication, the hypothesis testing amounts to distinguishing between the following states
\begin{align}
    \left\{\hat{\rho}_h^{\otimes n}\right\}_{h=1}^m\eqdef\left\{\left(\hat{\rho}_{\textnormal{th}}(N_{\textnormal{th}})^{\otimes n}\right)^{\otimes (h-1)}\otimes\hat{\rho}_{\textnormal{th}}(\kappa N+N_\textnormal{th})^{\otimes n}\otimes\left(\hat{\rho}_{\textnormal{th}}\left(N_\textnormal{th}\right)^{\otimes n}\right)^{\otimes (m-h)}\right\}_{h=1}^m.
\end{align}

By $M$-ary Quantum Chernoff bound, $E_{\textnormal{unassist}}$ is given as
\begin{align}
    E_{\textnormal{unassist}} &= \min_{(i,j):i\neq j}\max_{s\in[0,1]}-\log Q_s\left(\hat{\rho}_i\Vert \hat{\rho}_j\right)\\
    &= \max_{s\in[0,1]}-\log Q_s\left(\hat{\rho}_{\textnormal{th}}(N_{\textnormal{th}})\otimes\hat{\rho}_{\textnormal{th}}(\kappa N+N_\textnormal{th})\big\Vert\hat{\rho}_{\textnormal{th}}(\kappa N+N_\textnormal{th})\otimes\hat{\rho}_{\textnormal{th}}(N_\textnormal{th})\right)\\
    &= -2\log Q_{1/2}\left(\hat{\rho}_{\textnormal{th}}(\kappa N+N_\textnormal{th})\Vert\hat{\rho}_{\textnormal{th}}(N_\textnormal{th})\right)
\end{align}

Defining the geometric parameters of the thermal states as, let $r_1\eqdef\frac{\kappa N+N_\textnormal{th}}{\kappa N+N_\textnormal{th}+1}$ and $r_2\eqdef\frac{N_{\textnormal{th}}}{N_{\textnormal{th}}+1}$,
\begin{align}
    Q_{1/2}\left(\hat{\rho}_{\textnormal{th}}(\kappa N+N_{\textnormal{th}})\Vert \hat{\rho}_{\textnormal{th}}(N_{\textnormal{th}})\right) &= \sqrt{(1-r_1)(1-r_2)}\sum_{n=0}^\infty\left(\sqrt{r_1r_2}\right)^n\\
    &=\frac{\sqrt{(1-r_1)(1-r_2)}}{1-\sqrt{r_1r_2}}\\
    &= \frac{1}{\sqrt{(\kappa N+N_\textnormal{th}+1)(N_\textnormal{th}+1)}-\sqrt{(\kappa N+N_\textnormal{th})N_\textnormal{th}}}.
\end{align}

In the regime $N\ll1$ we obtain the approximation
\begin{align}
    E_{\textnormal{unassist},L}(N)= \frac{\kappa^2 N^2}{4N_\textnormal{th}(N_\textnormal{th}+1)}+\calO\left(N^3\right).
\end{align}
    \item \textit{Joint communication and Sensing Region:} For a fixed fraction $\lambda$ of the blocklength, the idler is shared with the receiver to enhance communication and while for the remaining fraction $1-\lambda$, the idler is retained to enhance sensing performance. In each case, the detection problem reduces to hypothesis testing with product states since we can eliminate the phase. We can consider the sharing of entanglement as a modulation scheme with fixed type $p_X$ on the alphabet $\calX=\{C,S\}$ in which $p_X(C)=\lambda,p_X(S)=\bar{\lambda}$. $C$ denotes entanglement is shared for assisted communication and $S$ denotes entanglement is kept for assisted sensing. Next, we define the states in mode $\hat{d}_h$ corresponding to the modulation scheme
    \begin{align}
        \hat{\sigma}^{C}_1 &\eqdef \hat{\rho}_{\textnormal{th}}\left(N_{\textnormal{th}}\right)\otimes\hat{\rho}_D,\\
        \hat{\sigma}^{C}_2 &\eqdef \hat{\rho}_D\otimes \hat{\rho}_{\textnormal{th}}\left(N_{\textnormal{th}}\right),\\
        \hat{\sigma}^{S}_1 &\eqdef \hat{\rho}_{TRI},\\
        \hat{\sigma}^{S}_2 &\eqdef \hat{\rho}_{RTI},.
    \end{align}
    Assuming that the controlled sensing exponent for quantum hypothesis testing derived in~\cite{Wang2022Joint} extends to infinite-dimensional quantum systems,
\begin{align}
    E &= \max_{s\in [0,1]}-\sum_{x\in\calX}p_X(x)Q_{s}\left(\hat{\sigma}_1^x\Vert\hat{\sigma}_2^x\right)\\
    &=\max_{s\in [0,1]} \left(-\lambda\log Q_s\left(\hat{\rho}_{\textnormal{th}}(N_{\textnormal{th}})\otimes\hat{\rho}_D\Vert\hat{\rho}_D\otimes\hat{\rho}_{\textnormal{th}}(N_{\textnormal{th}})\right)\right.\nonumber\\
    & \qquad\qquad\quad\qquad\left. -(1-\lambda)\log Q_s\left(\hat{\rho}_{TRI}\Vert\hat{\rho}_{RTI}\right)\right)\\
    &\overset{(a)}{=} \lambda E_{\textnormal{unassist}}(N)+(1-\lambda)E_{\textnormal{assist}}(N_S,N_{\textnormal{m}}),
\end{align}
where $(a)$ is because $s=0.5$ is optimal for both Chernoff bounds.

Since the protocol time-shares between the two schemes, the achievable communication rate is the corresponding convex combination of the rates in individual schemes,
\begin{align}
    R = \lambda R_{\textnormal{assist}}(N)+(1-\lambda) R_{\textnormal{unassist}}(N_S,N_\textnormal{m}).
\end{align}

Note that for a given $N$ such that the inequality in~\eqref{eq:rate_dpsk} is tight, for a fixed $N_S<N$ displaced \ac{PSK} rate is increasing in $N_\textnormal{m}$. Similarly, the quantum Chernoff bound in~\eqref{eq:QCB} $\overline{Q}_{1/2}$ only depends on the symplectic decompositions $(S_1,\{\nu_1,\nu_2\})$, resulting in no $N_\textnormal{m}$ dependence. Combining with the observation that the mean dependent part in~\eqref{eq:QCB} has a linear exponent in $N_\textnormal{m}$, we deduce that for a given $N_S$ choosing $N_\textnormal{m}=N-N_S$ is optimal.

Taking the union over $\lambda\in[0,1]$ and $N_S+N_{\textnormal{m}}\leq N$ completes the  of Theorem~\ref{thm:JBCS}.

Let uss focus on the regime $N\ll1$. Starting from the point $(N_S,N_\textnormal{m})=(N,0)$, shifting an amount $\Delta N= N\Delta\delta$ of mean photon number from $N_S$ to $N_\textnormal{m}$ results in a decrease in the exponent of order
\begin{align}
    \Delta E \eqdef E_{\textnormal{assist,L}}(N,\Delta\delta)-E_{\textnormal{assist,L}}(N,0)=\calO(N),
\end{align}
according to~\eqref{eq:exp_assist_l}, while the communication rate increases by
\begin{align}
    \Delta R = R_\textnormal{unassist,L}(N,\Delta\delta) = \calO(N),
\end{align}
according to~\eqref{eq:r_unassist_l}, corresponding to a slope of order $\calO(1)$. In contrast, the changes along time-sharing satisfies
\begin{align}
    \Delta R &= R_\textnormal{assist,L}(N,\Delta\delta) = \calO(-N\log N),\\
    \Delta E &= E_\textnormal{assist,L}(N,\Delta\delta)-E_\textnormal{assist,L}(N,0) = \calO(N),
\end{align}
according to~\eqref{eq:r_assist_L}, yielding a slope of order $\calO\left(-1/\log N\right)\leq\calO\left(1\right)$ as $N\to0$. Since, the slopes are actually negative this condition reduces to time-sharing slope to be larger than slope of the boundary of $(N_S,N\textnormal{m})$ region. This imbalance in the scalings of assisted and unassisted communication forces optimal power allocation to $(N_S,N_\textnormal{m})=(N,0)$, reducing to time-sharing.

\end{itemize}

\bibliographystyle{IEEEtran}
\bibliography{references}

@String{IEEE_ITW    = {Proc. of IEEE Information Theory Workshop}}

@ARTICLE{Bennett2002,
  author =       {Bennett, C.H. and Shor, P.W. and Smolin, J.A. and
                  Thapliyal, A.V.},
  journal =      {IEEE Transactions on Information Theory},
  title =        {Entanglement-assisted capacity of a quantum channel
                  and the reverse Shannon theorem},
  year =         2002,
  month =        {Oct},
  volume =       48,
  number =       10,
  pages =        {2637-2655},
  doi =          {10.1109/TIT.2002.802612}
}

@article{Wilde2017,
  title =        {Gaussian Hypothesis Testing and Quantum
                  Illumination},
  author =       {Wilde, Mark M. and Tomamichel, Marco and Lloyd, Seth
                  and Berta, Mario},
  journal =      {Phys. Rev. Lett.},
  volume =       119,
  issue =        12,
  pages =        120501,
  numpages =     6,
  year =         2017,
  month =        {Sep},
  publisher =    {American Physical Society},
  doi =          {10.1103/PhysRevLett.119.120501}
}

@article{Zhuang2021,
  title =        {Quantum Ranging with Gaussian Entanglement},
  author =       {Zhuang, Quntao},
  journal =      {Phys. Rev. Lett.},
  volume =       126,
  issue =        24,
  pages =        240501,
  numpages =     7,
  year =         2021,
  month =        {Jun},
  publisher =    {American Physical Society},
  doi =          {10.1103/PhysRevLett.126.240501}
}

@Article{He2023Integrated,
  author =       {He, Haijun and Jiang, Lin and Pan, Yan and Yi, Anlin
                  and Zou, Xihua and Pan, Wei and Willner, Alan E. and
                  Fan, Xinyu and He, Zuyuan and Yan, Lianshan},
  journal =      {Light: Science \& Applications},
  title =        {Integrated sensing and communication in an optical
                  fibre},
  year =         2023,
  month =        {Jan},
  issn =         {2047-7538},
  number =       1,
  pages =        25,
  volume =       12,
  doi =          {10.1038/s41377-022-01067-1},
  refid =        {He2023},
}

@Article{Yin2017,
  author =       {Yin, Juan and Cao, Yuan and Li, Yu-Huai and Liao,
                  Sheng-Kai and Zhang, Liang and Ren, Ji-Gang and Cai,
                  Wen-Qi and Liu, Wei-Yue and Li, Bo and Dai, Hui and
                  Li, Guang-Bing and Lu, Qi-Ming and Gong, Yun-Hong
                  and Xu, Yu and Li, Shuang-Lin and Li, Feng-Zhi and
                  Yin, Ya-Yun and Jiang, Zi-Qing and Li, Ming and Jia,
                  Jian-Jun and Ren, Ge and He, Dong and Zhou, Yi-Lin
                  and Zhang, Xiao-Xiang and Wang, Na and Chang, Xiang
                  and Zhu, Zhen-Cai and Liu, Nai-Le and Chen, Yu-Ao
                  and Lu, Chao-Yang and Shu, Rong and Peng, Cheng-Zhi
                  and Wang, Jian-Yu and Pan, Jian-Wei},
  journal =      {Science},
  title =        {Satellite-based entanglement distribution over 1200
                  kilometers},
  year =         2017,
  issn =         {0036-8075},
  month =        jun,
  number =       6343,
  pages =        {1140--1144},
  volume =       356,
  doi =          {10.1126/science.aan3211},
  eprint =
                  {http://science.sciencemag.org/content/356/6343/1140.full.pdf},
  publisher =    {American Association for the Advancement of Science},
}

@InProceedings{Wang2022Joint,
  author =       {Shi-Yuan Wang and Tuna Erdo\u{g}an and Uzi Pereg and
                  Matthieu R Bloch},
  booktitle =    IEEE_ITW,
  title =        {Joint Quantum Communication and Sensing},
  year =         2022,
  month =        aug,
  pages =        {506-511},
  doi =          {10.1109/ITW54588.2022.9965810},
  howpublished = {accepted to \emph{IEEE Information Theory Workshop}},
}

@Article{Li2016Discriminating,
  author =       {Ke Li},
  journal =      {The Annals of Statistics},
  title =        {Discriminating quantum states: The multiple Chernoff
                  distance},
  year =         2016,
  month =        {Aug},
  number =       4,
  volume =       44,
  doi =          {10.1214/16-aos1436},
  publisher =    {Institute of Mathematical Statistics},
}

@Article{Weedbrook2012,
  author =       {Weedbrook, Christian and Pirandola, Stefano and
                  Garc\'{\i}a-Patr\'on, Ra\'ul and Cerf, Nicolas
                  J. and Ralph, Timothy C. and Shapiro, Jeffrey H. and
                  Lloyd, Seth},
  journal =      {Rev. Mod. Phys.},
  title =        {Gaussian quantum information},
  year =         2012,
  month =        {May},
  pages =        {621--669},
  volume =       84,
  doi =          {10.1103/RevModPhys.84.621},
  issue =        2,
  numpages =     0,
  publisher =    {American Physical Society},
}

@article{Hiai2018,
    author = {Hiai, Fumio},
    title = "{Quantum f-divergences in von Neumann algebras. I. Standard f-divergences}",
    journal = {Journal of Mathematical Physics},
    volume = {59},
    number = {10},
    pages = {102202},
    year = {2018},
    month = {Sep},
    issn = {0022-2488},
    doi = {10.1063/1.5039973}
}

@article{Nair20,
author = {Ranjith Nair and Mile Gu},
journal = {Optica},
number = {7},
pages = {771--774},
publisher = {Optica Publishing Group},
title = {Fundamental limits of quantum illumination},
volume = {7},
month = {Jul},
year = {2020},
doi = {10.1364/OPTICA.391335},
}

@article{Pirandola2008,
  title = {Computable bounds for the discrimination of Gaussian states},
  author = {Pirandola, Stefano and Lloyd, Seth},
  journal = {Phys. Rev. A},
  volume = {78},
  issue = {1},
  pages = {012331},
  numpages = {8},
  year = {2008},
  month = {Jul},
  publisher = {American Physical Society}
}

@article{su2024,
  title={Achieving Quantum Advantage With Ring States Over Bosonic Channels},
  author={Shang-Jen Su and Shi-Yuan Wang and Matthieu R. Bloch and Zheshen Zhang},
  year={2026},
  month={May},
  journal={arXiv preprint},
  eprint={2410.17181},
  archivePrefix={arXiv},
  primaryClass={quant-ph},
  note={arXiv:2410.17181}
}

@InProceedings{MunarVallespir2024Joint,
  author =       {Munar-Vallespir, Pere and N\"otzel, Janis},
  booktitle =    {Proc. of IEEE 10th World Forum on Internet of Things
                  },
  title =        {Joint communication and sensing over the lossy
                  bosonic quantum channel},
  year =         2024,
  month =        {Nov},
  pages =        {1--6},
  doi =          {10.1109/wf-iot62078.2024.10811180},
}

@article{Audenaert2007,
  title = {Discriminating States: The Quantum Chernoff Bound},
  author = {Audenaert, K. M. R. and Calsamiglia, J. and Mu\~noz-Tapia, R. and Bagan, E. and Masanes, Ll. and Acin, A. and Verstraete, F.},
  journal = {Phys. Rev. Lett.},
  volume = {98},
  issue = {16},
  pages = {160501},
  numpages = {4},
  year = {2007},
  month = {Apr},
  publisher = {American Physical Society},
  doi = {10.1103/PhysRevLett.98.160501}
}

@ARTICLE{Chang2023,

  author={Chang, Meng-Che and Wang, Shi-Yuan and Erdoğan, Tuna and Bloch, Matthieu R.},

  journal={IEEE Journal on Selected Areas in Information Theory}, 

  title={Rate and Detection-Error Exponent Tradeoff for Joint Communication and Sensing of Fixed Channel States}, 
  month={May},
  year={2023},

  volume={4},

  number={},

  pages={245-259},

  doi={10.1109/JSAIT.2023.3275877}}

@INPROCEEDINGS{Chang2022,

  author={Chang, Meng-Che and Erdoğan, Tuna and Wang, Shi-Yuan and Bloch, Matthieu R.},

  booktitle={2022 2nd IEEE International Symposium on Joint Communications \& Sensing (JC\&S)}, 

  title={Rate and Detection Error-Exponent Tradeoffs of Joint Communication and Sensing}, 
  month={Mar},
  year={2022},

  volume={},

  number={},

  pages={1-6},

  doi={10.1109/JCS54387.2022.9743498}}

@ARTICLE{Joudeh2022,

  author={Joudeh, Hamdi and Willems, Frans M. J.},

  journal={IEEE Journal on Selected Areas in Information Theory}, 

  title={Joint Communication and Binary State Detection}, 
  month={Mar},
  year={2022},

  volume={3},

  number={1},

  pages={113-124},

  doi={10.1109/JSAIT.2022.3157999}}

@ARTICLE{Ahmadipour2024,

  author={Ahmadipour, Mehrasa and Kobayashi, Mari and Wigger, Michele and Caire, Giuseppe},

  journal={IEEE Transactions on Information Theory}, 

  title={An Information-Theoretic Approach to Joint Sensing and Communication}, 
  month = {Feb},
  year={2024},
  volume={70},
  number={2},
  pages={1124-1146},

  doi={10.1109/TIT.2022.3176139}}

@article{pant2019routing,
  title={Routing entanglement in the quantum internet},
  author={Pant, Mihir and Krovi, Hari and Towsley, Don and Tassiulas, Leandros and Jiang, Liang and Basu, Prithwish and Englund, Dirk and Guha, Saikat},
  journal={npj Quantum Information},
  volume={5},
  number={1},
  pages={25},
  month={Mar},
  year={2019},
  publisher={Nature Publishing Group UK London}
}

@article{tan2008QI,
  title = {Quantum Illumination with Gaussian States},
  author = {Tan, Si-Hui and Erkmen, Baris I. and Giovannetti, Vittorio and Guha, Saikat and Lloyd, Seth and Maccone, Lorenzo and Pirandola, Stefano and Shapiro, Jeffrey H.},
  journal = {Phys. Rev. Lett.},
  volume = {101},
  issue = {25},
  pages = {253601},
  numpages = {4},
  year = {2008},
  month = {Dec},
  publisher = {American Physical Society},
  doi = {10.1103/PhysRevLett.101.253601}
}

@article{Nussbaum2009Chernoff,
author = {Michael Nussbaum and Arleta Szkoła},
title = {{The Chernoff lower bound for symmetric quantum hypothesis testing}},
volume = {37},
journal = {The Annals of Statistics},
number = {2},
publisher = {Institute of Mathematical Statistics},
pages = {1040 -- 1057},
month = {Apr},
year = {2009},
doi = {10.1214/08-AOS593}
}

\end{document}